\documentstyle[twocolumn,aps]{revtex}
\begin{document} 
\twocolumn[\hsize\textwidth\columnwidth\hsize\csname @twocolumnfalse\endcsname
\draft
\title{Magnetic Impurities in Mott-Hubbard Antiferromagnets} 
\author{Avinash Singh$^{1,2}$ and Prasenjit Sen$^{2}$} 
\address{$^{1}$Theoretische Physik III, Universit\"{a}t Augsburg, 
86135 Augsburg, Germany \\
$^{2}$Department of Physics, Indian Institute of Technology, 
Kanpur 208016, India}
\maketitle
\begin{abstract}
A formalism is developed to treat magnetic impurities
in a Mott-Hubbard antiferromagnetic insulator within a representation
involving multiple orbitals per site. 
Impurity scattering of magnons is found to be strong when the number 
of orbitals ${\cal N}'$ on impurity sites is different from the number 
${\cal N}$ on host sites, leading to strong magnon damping and
singular correction to low-energy magnon modes in two dimensions.
The impurity-scattering-induced softening of magnon modes 
leads to enhancement in thermal excitation of magnons,
and hence to a lowering of the N\'{e}el temperature in
layered or three dimensional systems. 
Weak impurity scattering of magnons is obtained in the case
${\cal N}'={\cal N}$, where the impurity is represented in terms of modified 
hopping strength, and a momentum-independent multiplicative renormalization
of magnon energies is obtained, with the relative magnon damping decreasing 
to $q^2$ for long-wavelength modes. 
Split-off magnon modes are obtained when the impurity-host coupling is 
stronger, and implications are discussed for two-magnon Raman scattering.
The mapping between antiferromagnets and superconductors is utilized to
contrast formation of impurity-induced states.
\end{abstract} 
\vskip2pc] 
\section{Introduction}
While the problem of static impurities in antiferromagnetic
insulators is more than twenty five years old,\cite{cowley}
it has attracted renewed attention after the discovery of high-${\rm T_c}$
cuprate superconductors,\cite{bednorz} since their parent compounds
are antiferromagnetic insulators. From the very
early days of high-${\rm T_c}$ superconductivity a number of
doping studies have been done with various static impurities ---
both magnetic,\cite{xiao2} and nonmagnetic\cite{xiao,walstedt,mahajan}
--- replacing copper from the Cu-O planes as in La$_{2}$CuO$_{4}$.
Susceptibility measurements in high-${\rm T_c}$ cuprates doped with
magnetic impurities like Fe, Ni, Co
give evidence of local-moment formation,\cite{xiao} which
is expected to be intrinsically
associated with the magnetic impurities. This is unlike the case of
nonmagnetic impurities such as Zn, Al, Ga etc. which,
despite being intrinsically nonmagnetic, give rise to local moments
in the copper-oxide planes when doped in
cuprate antiferromagnets. This was inferred earlier from the
Curie-Weiss behavior of the magnetic
susceptibility,\cite{xiao,gee} and has been recently confirmed
in the Y-NMR studies of doped  1-2-3 systems as seen in the
progressively increasing linewidth of the Y-NMR signal
with decreasing temperature.\cite{mahajan,alloul}
Xiao {\em et al.} have also ascertained the spin states of different
magnetic dopants from the observed local moments,\cite{xiao2}
and find, for example,  that Fe is in a spin-$\frac{5}{2}$ state, whereas Ni
is in a spin-1 state.
They also find a correlation between ${\rm T_c}$ reduction and
size of the local moment, consistent with the magnetic pair breaking
mechanism.

Although theoretically the problem of magnetic
impurities in an antiferromagnet has been studied recently 
within the Heisenberg representation of localized spins,\cite{dkumar}
no such comprehensive study exists within the Mott-Hubbard model,
which provides a good description of the 3d holes in the Cu-O planes
of high-${\rm T_c}$ superconductors. 
Recently the problem of nonmagnetic impurities in the Mott-Hubbard 
antiferromagnet was addressed and 
defect states, local-moment formation, impurity-scattering of magnons,
and finite-temperature magnetic dynamics in layered systems were 
studied.\cite{gapstate,swscattering}
Other recent works on static vacancies in antiferromagnets include
exact diagonalization studies with Heisenberg model,\cite{bulut}
linear spin wave theory,\cite{brenig+kampf} and 
exact diagonalization of strongly correlated small clusters.\cite{poilblanc} 
While nonmagnetic impurities can be simply represented by
spin-independent impurity potential, the situation
is more complex for magnetic impurities. 
In this paper we introduce several representations to treat 
magnetic impurities in different situations.
A simple extension to spin-dependent impurity potential is 
followed by a more sophisticated approach involving a generalized ${\cal N}$-orbital
Hubbard model with multiple orbitals per site.
Broadly there are two distinct classes depending
on whether the number of orbitals ${\cal N}'$ at the impurity site
is the same as or different from the number  of orbitals ${\cal N}$
at the host sites. In the case ${\cal N}'={\cal N}$ the magnetic impurity
is represented through a modified hopping strength $t'$
between the impurity orbitals and the neighboring host orbitals.
In the strong-correlation limit ($U \gg t$) wherein the Mott-Hubbard AF
with ${\cal N}$ orbitals per site maps to the spin $S={\cal N}/2$
quantum Heisenberg AF, the modified hopping strength translates into
modified exchange coupling $J'=4t'^{2}/U$ between the impurity spin
and the neighboring host spins. This describes the situation
where, in spin language, the impurity spin $S'$ is equal to the host
spin $S$, but is coupled to its neighbors with a different exchange interaction  $J'$.
Similarly the case ${\cal N}' \ne {\cal N}$ with no modification in
hopping strength or Hubbard interaction energy corresponds
to the situation where the impurity spin is different from the host spins ($S'\ne S$).

\section{Single-orbital magnetic impurity}
In this section we consider a single-orbital magnetic impurity
embedded in an AF host which is described by the Hubbard model
with one orbital per site with exactly half filling.
For concreteness we consider the square lattice, generalization to
other bipartite lattices being straightforward. The host
Hamiltonian is
  \begin{equation}
  H_{0}=-t\sum_{\langle ij \rangle \sigma}
  (a^{\dagger}_{i \sigma}a_{j  \sigma} +  
  a^{\dagger}_{j \sigma}a_{i  \sigma} ) 
  + U\sum_i n_{i \uparrow}n_{i \downarrow},
      \label{eq:ham}
  \end{equation}
where $t$ is the nearest-neighbor (NN) hopping strength and $U$ 
the on-site Coulomb repulsion. The AF state and its
associated features such as sublattice magnetization, magnon energies,
quantum corrections etc. have been studied earlier in detail.\cite{quantum} 
We model the single-orbital impurity in terms of locally modified hopping
term $t'$ between the impurity orbital and its NN host orbitals.
The Hamiltonian with such an impurity on site $I$ can be written as
below, where the sum is over all nearest neighbors $J$ of
the impurity sites $I$, and $\delta t=t' -t$ is 
the hopping perturbation around the impurity site,
\begin{equation}
  H = H_0 + \delta t \sum_{\langle IJ \rangle \sigma}
  (a^{\dagger}_{I \sigma}a_{J \sigma} +
   a^{\dagger}_{J \sigma}a_{I \sigma} )
  \end{equation}
  
We start with the perturbative method where the impurity-induced perturbation
$[\delta\chi^{0}]\equiv [\chi^0]-[\chi^0_{\rm host}]$ 
to the zeroth-order, antiparallel-spin,
particle-hole propagator is obtained in powers of $\delta t/t$, 
and resulting corrections to its eigenvalues  
then yield the renormalization in magnon energies.\cite{hopping}
Diagrammatic contributions to
$[\delta\chi^{0}]$ to first order in $\delta  t$, and their evaluation
in the strong-correlation limit have been discussed 
earlier in context of the hopping disorder problem.\cite{hopping}
We obtain for the diagonal, off-diagonal, and nearest-neighbor diagonal
terms, expressed in units of $-t^2/\Delta^3$ for convenience, 
\begin{equation}
[\delta \chi^{0}]_{II} =\frac{z}{2}\frac{\delta t}{t}
 ; \;\;\;\;\;
[\delta \chi^{0}]_{IJ} = [\delta \chi^{0}]_{JI}
= [\delta \chi^{0}]_{JJ}= \frac{1}{2}\frac{\delta t}{t}
\end{equation}
where $z=4$ is the coordination number for the square lattice,
$2\Delta \approx U$ is the Hubbard gap, and only terms upto
order $(t^{2}/\Delta^{3})$ have been retained, appropriate  to  the
strong-correlation  limit. We  notice  that the sum of all
matrix elements diagonal in sublattice basis, $[\delta \chi^{0}]_{II}+
[\delta \chi^{0}]_{JJ}$ is
precisely equal to the sum of off-diagonal  matrix  elements $[\delta   \chi^{0}]_{IJ}
+[\delta \chi^{0}]_{JI}$.
An immediate  consequence  of this  correlation  is that the
Goldstone   mode  is  preserved  and  that   generally  the  effective
scattering of low-energy, long-wavelength magnon modes is weak.

If the impurity is on an A-sublattice site, then
for the first-order correction we obtain after summing over nearest 
neighbor terms, 
\begin{eqnarray}
&\ & \delta\lambda_q ^{(1)}\equiv 
\langle q | [\delta \chi^0] | q \rangle = \nonumber \\
 && \alpha^2 [\delta\chi^{0}]_{II} +
\alpha\beta z\gamma_q[\delta \chi^{0}]_{IJ} 
 +\beta\alpha z\gamma_q[\delta \chi^{0}]_{JI}
+\beta^2 z[\delta\chi^{0}]_{JJ} 
\end{eqnarray}
where $\alpha$ and $\beta$ are the magnon amplitudes on A and
B sublattices respectively, and $\gamma_q =(\cos q_x +\cos q_y )/2$.
An identical result is obtained
when the impurity is on a B-sublattice site, because in this case 
$\alpha$ and $\beta$ are simply exchanged in the above equation, and since
$[\delta\chi^0]_{II}=z[\delta\chi^0]_{JJ}$, this expression is symmetric
under exchange of $\alpha$ and $\beta$. 
Using $\alpha=\sqrt{\frac{1}{N}(1-\omega_q ^0)}$ and 
$\beta=-\sqrt{\frac{1}{N}(1+\omega_q ^0)}$, 
where $\omega_q ^0 = \sqrt{1-\gamma_q ^2}$ is the 
host magnon energy in units of $2J$ for the momentum-$q$ mode, we obtain
after summing over contributions from all impurities
\begin{equation}
\delta \lambda_q ^{(1)} = x 
z\frac{\delta t}{t}(1-\gamma_q ^2), 
\end{equation}
where $x$ is the total impurity concentration, 
and impurities are assumed to be evenly distributed between the two sublattices. 
The renormalized magnon energy, given by the pole in the magnon propagator, 
is now obtained from the solution of the equation
$1- \sqrt{\omega^2 +\gamma_q ^2} + \delta \lambda_q ^{(1)} = 0$, and 
upto first order in the effective impurity strength $x \delta t/t$  we obtain
\begin{equation}
\omega_{q}=\omega_{q}^{0}\left (1+x\, z\frac{\delta t}{t}\right ) .
\end{equation}
This result agrees exactly with
the calculations\cite{dkumar} on the Heisenberg model in that
there are no singular corrections to the magnon energy in
the case $S^{\prime} = S$, and the correction is proportional to
$x \delta t/t = (1/2) x \delta J /J$. 

Turning now to the magnon-energy renormalization of the 
localized, high-energy modes with energy near $2J$, which correspond to
local spin deviation, we have  $\alpha=0$, $\beta=1$, so that
$\delta \lambda ^{(1)} = \frac{1}{2} z (\delta t/t)$. This implies that
the magnon energy gets shifted from $2J$ to
\begin{equation}
\omega=2J\left (1+\frac {z}{2}\frac{\delta t}{t}\right ) .
\end{equation}
In this case the impurity concentration does not enter
the magnon-energy renormalization, rather it has a bearing on the
spectral weight of these high-energy modes. Thus for $\delta t$ positive,
the magnon spectrum goes up by energy $zJ\delta t /t$.
This increase is expected from the simple picture of these high-energy modes
corresponding to local spin deviations. The energy cost of making a spin
deviation on the impurity site is $zJ'/2$, where $J'/2$ is the bond strength.
With $t'=t(1+\delta t/t)$, to first order in $\delta t/t$ we have
$\Delta \epsilon=z(J'-J)/2=zJ\delta t /t$.

The exact-eigenstates analysis also shows that precisely one magnon
state at the upper end of the spectrum is split off from the
magnon energy band. These split-off modes are strongly localized around
the impurity sites, and hence correspond to local spin deviations.
Furthermore, for different values of the impurity hopping $t'/t$ 
it is seen from the magnon spectrum that the energy separation 
of the split-off state from the upper end of the spectrum
increases roughly in proportion to $\delta t$, as obtained in the
perturbative analysis. 
This exact-eigenstates approach for obtaining magnon energies
and wavefunctions from the fermionic eigensolutions in the self-consistent 
AF state has been described earlier.\cite{selfcon}

\section{Spin-dependent impurity potential}
Nonmagnetic impurities in the Mott-Hubbard
AF were modelled earlier via a spin-independent impurity potential
term, and as a natural extension we therefore consider
the following spin-dependent impurity term for magnetic impurities,
\begin{equation}
H_{\rm imp}^{\rm mag}=\sum_{I}\Psi_{I}^{\dagger}[-\sigma_3 V ] 
\Psi_{I} ,
\end{equation}
where $\Psi_{I}=(a_{I\uparrow}\;\;a_{I\downarrow})$. A spin-independent
impurity potential $\epsilon_0$ can be included for generality,
however, we shall consider the limit $V>>\epsilon_0$, so that
the potential for spin $\sigma$ fermion is $V_{\sigma} \approx -\sigma V$.
We choose $V$ to be positive for impurities on the A-sublattice sites,
so that $V_{\uparrow}$ is very low and $V_{\downarrow}$ is very high.
The sign of $V$ is reversed for impurities on B-sublattice sites.
This choice of potential ensures that the magnetization on the impurity
sites follows the host AF ordering. Such a spin-dependent impurity
potential can arise from a coupling $-\vec{\sigma}. \vec{S}_{\rm imp}$
between the itinerant fermion spin $\vec{\sigma}$ and the
static magnetic impurity spin
$\vec{S}_{\rm imp}$, resulting from a strong Hubbard interaction. 
Since experiments on high-${\rm T_c}$ cuprates show the impurity spin to be
antiferromagnetically coupled with the host spins,\cite{xiao2}
we take the local
field direction to be along the local magnetization direction ($\hat{z}$).
The low potential (for spin-up)
is justified in view of the fact that the ionization energy
for both ${\rm Fe^{+3}}$ and ${\rm Ni^{+2}}$ {\it
i.e.,} the fourth and the third ionization energies
respectively for Fe and Ni are much higher than the third
ionization energy for Cu. Whereas the ionization energies for
${\rm Fe^{+3}}$ and ${\rm Ni^{+2}}$ are 54.8 eV and 35.17 eV
respectively, the ionization energy for ${\rm Cu^{+2}}$ is 20.2 eV.

We now examine formation of impurity-induced states due to this spin-dependent impurity potential.
Within the T-matrix analysis, used earlier for nonmagnetic impurities,\cite{gapstate}
energies of impurity-induced states are obtained from 
solutions of $g_{II}^{\sigma}(\omega)=1/V_\sigma$. 
For large $|V|/U$ these impurity states are formed at energies $\sim -\sigma V$ 
for the two spins, and are essentially
site localized and therefore decoupled from the system. 
Thus, for the magnetic-impurity case when the 
impurity spin is antiferromagnetically coupled to the neighboring
host spins, a significant difference from the nonmagnetic-impurity case is that
there are no defect states formed in the Hubbard gap.
Rather only impurity states are formed, far removed in energy from the Hubbard bands.

Within the above representation of magnetic impurities in terms
of spin-dependent impurity potential, 
the fermion number is unchanged, unlike the case of nonmagnetic
impurities where one fermion is removed for every added impurity.
Hence the impurity sites do not quite act as spin vacancies.
Nonetheless, the presence of a impurity potential term which breaks time-reversal
symmetry leads to a partial decoupling of the impurity site from the host. 
This is most easily seen in the limit $V\rightarrow\infty$ 
where the local antiparallel-spin, particle-hole excitations are
suppressed by the large energy difference $2V$, leading to an absence of the 
$\omega$ term, and therefore to strong magnon scattering. 
Quite generally, the particle-hole energy difference for antiparallel spins
is modified by the spin-dependent impurity potential from $2\Delta$ to $2\Delta + 2V $,
leading to a modification in the $\omega $ term. 
For spin-independent impurity potential the particle-hole energies are
shifted equally, and hence it is the removal of a fermion from the
impurity site that is crucial. 
As a result of this decoupling of magnetic impurity sites, 
a qualitatively identical
impurity-induced perturbation $[\delta\chi^0(\omega)]$ is obtained,
leading to similar results for magnon renormalization as for the 
nonmagnetic-impurity case, where singular corrections were obtained
for low-energy magnon modes in two dimensions.\cite{swscattering}
The strong impurity-scattering
of magnons also introduces significant damping, with the ratio of
magnon damping term to its energy being simply proportional to the impurity 
concentration $x$ for long-wavelength modes.

\section{Generalized Hubbard-model representation}
In order to represent higher-spin magnetic impurities, 
we now generalize to the situation with ${\cal N}$ and ${\cal N}'$ orbitals 
on host and impurity sites respectively.
An appropriate model for this case is the generalized
Hubbard model with multiple orbitals per site. This model has been used
earlier to study quantum corrections in the antiferromagnetic state
in a spin-rotationally-symmetric formalism, where a systematic perturbative
expansion in powers of $1/{\cal N}$ was developed.\cite{quantum}
We introduce a slight extension here in this model which makes it
equivalent, in the strong correlation limit, to the spin-$S$
QHAF, where $S={\cal N}/2$. The modification is to allow the NN hopping
term to operate between {\em all} orbitals, whereas the hopping term
considered earlier was diagonal in the orbital index.\cite{quantum}
We therefore consider the following Hamiltonian for the AF host,
\begin{eqnarray}
H&=&-t\sum_{<ij>\sigma\alpha\beta}( a_{i\sigma\alpha}^{\dagger}
a_{j\sigma\beta}+{\rm h.c.}) \nonumber \\
&+&\frac{U}{\cal N}\, \sum_{i\alpha\beta}
(a_{i\uparrow\alpha}^{\dagger}a_{i\uparrow\alpha}
a_{i\downarrow\beta}^{\dagger}a_{i\downarrow\beta} +
a_{i\uparrow\alpha}^{\dagger}a_{i\uparrow\beta}
a_{i\downarrow\beta}^{\dagger}a_{i\downarrow\alpha} )
\end{eqnarray}
where $\alpha$ and $\beta$ are the orbital indices which run
from 1 to ${\cal N}$, and the two Hubbard interaction terms
are respectively direct and exchange type interactions with respect
to orbital indices.
In the symmetric case when the two interaction strengths are
identical, as considered here, the system possesses
spin-rotational symmetry. 
It has been shown earlier that in the symmetric case
the two interaction terms can together be written as
$H_{\rm int}=-(U/{\cal N})
\sum_{i}(\vec{S}_{i}. \vec{S}_{i} + n_{i}^{2}) $,
where $\vec{S}_i$ and $n_i$
are the total spin and charge density operators, respectively.
Spin-rotational-symmetry is therefore inherent in this
impurity representation as well. Furthermore, in the strong correlation
limit, a strong Hund's coupling exists which energetically favors
the maximum multiplicity case ($S={\cal N}/2$) for
the total spin operator $\vec{S}_i$. 

Magnetic impurities are represented by
introducing ${\cal N}'\ne {\cal N}$ orbitals at the impurity sites.
We first examine the transverse spin fluctuation 
propagator in the host AF state,
$\chi ^{-+}(rt,r't')\equiv \langle \Psi | S^{-}(rt) S^{+}(r't') |\Psi
\rangle$, where $\vec{S}(rt)=\sum_{\alpha}\psi_{\alpha}^{\dagger}(rt)
\frac{\vec{\sigma}}{2}\psi_{\alpha}(rt)$ is the {\em total}
spin  operator.
Again, at the RPA level the magnon propagator is given by
$\chi^{0}(\omega)/[{\bf 1}-(U/{\cal N})\chi^{0}(\omega)]$,
where $[\chi^{0}(\omega)]$ now involves orbital summations,
with matrix elements given by,
\begin{equation}
[\chi^{0}(\omega)]_{ij}=i\int\frac{d\omega'}{2\pi}
\sum_{\alpha\beta}
G^{\uparrow}_{i\alpha,j\beta}(\omega')
G^{\downarrow}_{j\beta,i\alpha}(\omega'-\omega).
\end{equation}

Since each orbital is now connected via hopping to $\cal N$ orbitals
on the NN sites, the electronic spectral weights are correspondingly
modified. For example, in the strong-correlation limit,
the on-site majority and minority spin densities in each orbital are
now $1-{\cal N}t^2/\Delta^2$ and ${\cal N}t^2/\Delta^2$
respectively.
A straightforward extension of the earlier analysis in the
strong-correlation limit\cite{quantum} leads to:
\begin{equation}
\chi^{0}(q,\omega)= {\cal N}\frac{1}{U}{\bf 1}-{\cal N}^2
\frac{D}{2}\frac{t^{2}} {\Delta^{3}}
\left [\begin{array}{cc}
1+\frac{\omega}{DJ{\cal N}}  & \gamma_{q}  \\
\gamma_{q}         & 1-\frac{\omega}{DJ{\cal N}}
\end{array}\right ]
\end{equation}
where $J=4t^2/U$ as usual, and $D$ is dimensionality of the hypercubic
lattice.
Since different orbitals on the same site are not directly coupled,
the intrasite propagator is diagonal in orbital index, and therefore
the leading order diagonal terms (the $1/U$ and the $\omega$ term)
are proportional to ${\cal N}$. However, the NN hopping operates between
all orbitals, and therefore the off-diagonal term and the next-to-leading
order piece (arising from hopping) in the diagonal term
are both proportional to ${\cal N}^2$.
The magnon energies are now given by: $\omega_q=
DJ{\cal N} \sqrt{1-\gamma_q ^2}=2DJS\sqrt{1-\gamma_q ^2}$
in terms of the spin $S={\cal N}/2$.

We now introduce a magnetic impurity in the system with
${\cal N}'\ne {\cal N}$ orbitals at the impurity site $I$.
The resulting modification in the electronic spectral weights
leads to the following changes in the $[\chi^{0}(\omega)]_{ij}$
matrix elements for $i,j$ in the vicinity of the impurity site $I$ :
\begin{eqnarray}
\ [\chi^{0}]_{II} &=& {\cal N}'\frac{1}{U} - {\cal N}{\cal N}'
\frac{D}{2}
\frac{t^2}{\Delta^3}\left ( 1+\frac{\omega}{DJ{\cal N}} \right )\nonumber \\
\ [\chi^{0}]_{IJ} &=& -{\cal N}{\cal N}'\frac{D}{2}
\frac{t^2}{\Delta^3} \frac{1}{z} \nonumber \\
\ [\chi^{0}]_{JJ} &=& {\cal N}\frac{1}{U} -{\cal N}^2
\frac{D}{2}\frac{t^2}{\Delta^3}\left ( 1-\frac{\omega}{DJ{\cal N}}
\right ) \nonumber \\
  &-& {\cal N}({\cal N}'-{\cal N})\frac{D}{2}\frac{t^2}{\Delta^3}
\frac{1}{z} 
\end{eqnarray}

Since now the local Hubbard interaction strength itself is not uniform but
depends on the number of site orbitals, we have to multiply the
$[\chi^0]$ matrix with the diagonal interaction matrix $[{\cal U}]$
containing elements $U/{\cal N}$ for host sites and $U/{\cal N}'$
for the impurity site. We therefore examine the local matrix elements
of the matrix product $[{\cal U}\chi^0]_{ij}={\cal U}_{ii}\chi^{0}_{ij}$
for $i,j$ in the vicinity of the impurity site.
The impurity-induced perturbation in the matrix elements of the
product $[{\cal U}\chi^0]$ are obtained as below:
\begin{eqnarray}
& \delta[{\cal U}\chi^0(\omega)]_{II} \;\; = & 0 \nonumber \\
& \delta[{\cal U}\chi^0(\omega)]_{IJ} \;\; = & 0 \nonumber \\
& \delta[{\cal U}\chi^0(\omega)]_{JI} \;\; = & -U ({\cal N}'-{\cal N})
\frac{D}{2}\frac{t^2}{\Delta^3} \frac{1}{z}\nonumber \\
& \delta[{\cal U}\chi^0(\omega)]_{JJ} \;\; = &
-U ({\cal N}'-{\cal N})
\frac{D}{2}\frac{t^2}{\Delta^3}\frac{1}{z} .
\end{eqnarray}

We now obtain the magnon-energy renormalization by
perturbatively obtaining the impurity-induced correction
to the eigenvalues of the $[{\cal U}\chi^0(\omega)]$ matrix.
As discussed earlier,\cite{swscattering} we 
treat $\delta[{\cal U}\chi^{0}(\omega)]$ as the perturbation matrix,
and determine corrections to eigenvalues of
$[\chi^{0}_{\rm host}(\omega)]$.
Evaluating the first-order correction $\langle q | \delta [
{\cal U}\chi^0 (\omega)]| q \rangle $ from the magnon eigenvector $|q\rangle$,
and retaining terms to first order only, 
we obtain: 
\begin{equation}
\delta \lambda_q ^{(1)}=U ({\cal N}'-{\cal N})
\frac{D}{2}\frac{t^2}{\Delta^3}\frac{\omega}{DJ}.
\end{equation}
As for the nonmagnetic impurity case, we obtain here
a correction which is linear in energy, and this 
signifies strong impurity scattering of magnons for long-wavelength, low-energy
modes, leading to singular corrections in two dimensions and strong magnon damping 
from second-order scattering processes.\cite{swscattering} 
The scattering
term is explicitly proportional to the difference $({\cal N}'-{\cal N})$
between the number of orbitals on the impurity site and the host sites,
which arises from the different dynamics of the impurity spin and the host
spins. This generally implies that impurity scattering of magnons is strong
when the impurity spin $S'={\cal N}'/2$ is different from the host spin
$S={\cal N}/2$, in agreement with earlier studies within the Heisenberg model,\cite{dkumar} 
and the one-band model with nonmagnetic impurities where ${\cal N}=1$ 
and ${\cal N}'=0$.\cite{swscattering}

\section{Conclusions}
In conclusion, we have developed a formalism to treat magnetic impurities
in a Mott-Hubbard antiferromagnetic insulator within a representation
involving multiple orbitals per site. 
For the case ${\cal N}'={\cal N}$, when the impurity spin is identical to
the host spin, the magnetic impurity is
represented by locally modified hopping strength, and 
we find that the effective scattering of long-wavelength magnon modes
is weak, leading to momentum-independent multiplicative renormalization
of magnon energies. For positive hopping perturbation $\delta t$
we find localized, split-off magnon modes corresponding to local
spin deviations at impurity sites. These split-off modes will be relevant 
in two-magnon Raman scattering which probes high energy magnetic excitations.
In the other case ${\cal N}'\ne {\cal N}$, when the impurity spin 
is different from the host spin,
we obtain strong impurity scattering of magnon modes proportional
to the difference (${\cal N}' - {\cal N}$), leading to
singular corrections in two dimensions and strong magnon damping.
The impurity-scattering-induced softening of magnon modes 
implies enhancement in thermal excitation of magnons,
and hence to a lowering of the N\'{e}el temperature in
layered or three dimensional systems.  We also find that the
process of putting additional impurity orbitals leads to enhanced
impurity magnetization and localization of electronic states at the
impurity, indicating partial decoupling of the impurity site from the  host.
A unique feature of having multiple impurity orbitals is
the presence of exactly site-localized eigenstates in the electron
spectrum which are completely antisymmetric between impurity orbitals.

When the magnetic impurity is represented in terms of a spin-dependent
impurity potential, we find that 
the breaking of time-reversal symmetry leads to a decoupling of the 
impurity site from the host, and strong magnon scattering similar to 
the case of spin vacancies is obtained. 
We also find that when the magnetic impurity spin is antiferromagnetically
coupled to the neighboring host spins, only impurity states are formed, and
there are no defect states formed within the Hubbard gap. 
The local moment associated with
the magnetic impurity therefore intrinsically arises from the
spin-density difference at the impurity site. 
Using the well-known particle-hole transformation, 
The problem of magnetic impurities in an AF can be mapped
to that of nonmagnetic impurities in a superconductor,
which is characterized by absence of defect states within the 
superconducting gap and robustness of superconducting gap.\cite{supercond} 
Conversely, a nonmagnetic impurity in
a positive-$U$ Hubard AF maps onto a magnetic impurity in a negative-$U$ Hubbard superconductor,
and here defect states are formed within the gap in both cases.\cite{shiba,maki}

\newpage
\section*{Acknowledgments}
Helpful conversations with S. N. Basu, S. Tewari, D. Sa, 
V. Subrahmanyam, and V. A. Singh are gratefully acknowledged. 
This work was supported in part by a Research Grant (No. SP/S2/M-25/95)
from the Department of Science and Technology, India. 
A.S. also acknowledges support from the Alexander von Humboldt Foundation.


\newpage
\begin{references}
\bibitem[1]{cowley}
For a review, see R. A. Cowley and W. J. L. Buyers, Rev. Mod. Phys.
{\bf 44}, 406 (1972). 
\bibitem[2]{bednorz}
J. G. Bednorz and K. A. M\"{u}ller, Z. Phys. B {\bf 64}, 188, (1986).
\bibitem[3]{xiao2}
G. Xiao, M. Z. Cieplak, J. Q. Xiao, and C. L. Chien, Phys. Rev. B
{\bf 42}, 8752 (1990).
\bibitem[4]{xiao}
G. Xiao, M. Z. Cieplak, A. Gavrin, F. H. Streitz, A. Bakhshai and
C. L. Chien, Phys. Rev. Lett, {\bf 60}, 1446 (1988).
\bibitem[5]{walstedt}
R. E. Walstedt, R. F. Bell, L. F. Schneemeyer, and T. V. Waszcazk,
Phys. Rev. B {\bf 48}, 10 646 (1993).
\bibitem[6]{mahajan}
A. V. Mahajan, H. Alloul, G. Collin, and J. F. Marucco, Phys. Rev. Lett.
{\bf 72}, 3100 (1994).
\bibitem[7]{gee}
C. -S. Gee {\em et al.}, J. Superconductivity, {\bf 1}, 63, (1988).
\bibitem[8]{alloul}
H. Alloul, P. Mendels, H. Casalta, J. F. Marucco, and J. Arabski,
Phys. Rev. Lett. {\bf 67}, 3140 (1991).
\bibitem[9]{dkumar}
C. C. Wan, A. B. Harris and D. Kumar, Phys. Rev. B {\bf 48}, 1036 (1993).
\bibitem[10]{gapstate}
P. Sen, S. Basu and A. Singh, Phys. Rev. B {\bf 50}, (RC) 10381 (1994).
\bibitem[11]{swscattering}
P. Sen and A. Singh, Phys. Rev. B {\bf 53}, 328 (1996).
\bibitem[12]{bulut}
N. Bulut. D. Hone, D. J. Scalapino, and E. Y. Loh, Phys. Rev. Lett. {\bf 62}, 2192 (1989).
\bibitem[13]{brenig+kampf}
W. Brenig and A. Kampf, Phys. Rev. B {\bf 43}, 12 914 (1991).
\bibitem[14]{poilblanc}
D. Poilblanc, D. J. Scalapino, and W. Hanke, Phys. Rev. Lett. {\bf 72}, 884 (1994). 
\bibitem[15]{quantum}
A. Singh, Phys. Rev. B {\bf 43}, 3617 (1991).
\bibitem[16]{hopping}
S. Basu and A. Singh, Phys. Rev. B {\bf 55}, 12 338 (1997). 
\bibitem[17]{selfcon}
S. Basu and A. Singh, Phys. Rev. B {\bf 53}, 6406 (1996).
\bibitem[18]{two-magnon}
S. Basu and A. Singh, Phys. Rev. B {\bf 54}, 6356 (1996).
\bibitem[19]{supercond}
P. W. Anderson, J. Phys. Chem. Solids {\bf 11}, 26 (1959).
\bibitem[20]{shiba}
H. Shiba, Prog. Theor. Phys. {\bf 40}, 435 (1968).
\bibitem[21]{maki}
K. Maki, {\it Superconductivity}, Vol II, edited by R. D. Parks (Dekker, New York, 1969). 
\end{references}
\end{document}